\documentclass[superscriptaddress,twocolumn,amsmath,amssymb,footinbib]{revtex4-2}
\usepackage[english]{babel}
\usepackage[pdftex]{graphicx}
\usepackage{gensymb}
\usepackage[symbol]{footmisc}
\usepackage{xcolor}
\usepackage{lineno}
\usepackage{csquotes}
\usepackage{mathrsfs}
\usepackage{subcaption}
\usepackage[colorlinks=true, linkcolor=blue, citecolor=blue, urlcolor=blue]{hyperref}
\hypersetup{
    colorlinks=true,
    linkcolor=blue,
    filecolor=magenta,      
    urlcolor=blue,
    pdftitle={Manuscript},
    pdfpagemode=FullScreen,
}

\usepackage[capitalize]{cleveref}
\crefformat{equation}{Eq.~(#2#1#3)} 

\usepackage{caption} 
\usepackage{ragged2e}
\DeclareCaptionJustification{myjustified}{\justifying}
\begin{document}

\newcommand{\physicsDept}{Department of Physics, New Mexico State University, Las Cruces, NM 88001, USA}
\newcommand{\CHMEDept}{Department of Chemical and Materials Engineering, New Mexico State University, Las Cruces, NM 88001, USA}

\newcommand{\ang}{\text{\normalfont\AA}}
\newcommand{\dpar}{$d_{||}$ }
\newcommand{\dxzyz}{$(d_{\text{xz}}$, $d_{\text{yz}})$ }
\renewcommand{\thefootnote}{\fnsymbol{footnote}}
\newcommand{\red}{\textcolor{red}}
\newcommand{\blue}{\textcolor{blue}}

\title{Enhanced-Entropy Phases in Geometrically Frustrated Pyrochlore Magnets}

\author{Prakash Timsina}
\affiliation{\physicsDept}
\author{Andres Chappa}
\affiliation{\physicsDept}
\author{Deema Alyones}
\affiliation{\CHMEDept}
\affiliation{\physicsDept}
\author{Igor Vasiliev}
\affiliation{\physicsDept}
\author{Ludi Miao*}
\affiliation{\physicsDept}

\date{\today}


\begin{abstract}
\begin{center}
$^{*}$ Email: lmiao@nmsu.edu
\end{center}
Frustrated magnets provide a platform for exploring exotic phases beyond conventional ordering, with potential relevance to functional materials and information technologies. In this work, we use Monte Carlo simulations to map the thermodynamic phase diagram of pyrochlore iridates $R_2$Ir$_2$O$_7$ ($R$ = Dy, Ho) with three stable magnetic ground-state phases: frustrated spin-ice 2-in–2-out (2I2O) phase, frustrated fragmented 3-in–1-out/1-in–3-out (3I1O/1I3O) phase, and antiferromagnetic all-in–all-out (AIAO) phase without frustration. We investigate the physical origin of two enhanced-entropy (EE) phases at finite temperatures, which emerge at the boundaries between stable phases. These EE phases exhibit high magnetic susceptibility and high entropy without long-range order. Their stabilization arises from entropic minimization of the free energy, where the entropy dominates energetic competition near phase boundaries at finite temperatures. Our results demonstrate a platform to engineer highly susceptible and degenerated states through frustration and thermal activation, offering a foundation for the design of entropy-based materials and phases in correlated systems.
\end{abstract}

\maketitle
\section*{Introduction}
Entropy, which arises from extensive configurational degeneracy, is pivotal in stabilizing exotic phases and emergent functionalities, spanning from chemically disordered high-entropy alloys \cite{George2019, Ye2016, Miracle2019, Rost2015, Bu2024, Ouyang2024, Zhang2017, Odetola2024, Wu2020} and entropy-stabilized ceramics \cite{Oses2020,Divilov2024} to geometrically frustrated states such as quantum spin liquids \cite{Wen2019,Broholm2020,Takayama2025,Gingras2014}, spin ice \cite{Ramirez1999,Snyder2001,Morris2009}, and frustrated pyrochlore magnets \cite{Tokiwa2016,Harris1997,Vayer2021,Vayer2023,Vayer2024,Lacroix2011,Gardner2010}. Among these examples, pyrochlore magnets stand out for highly tunable magnetic structures and degrees of frustration, arising from a delicate balance of competing interactions \cite{Lefrançois2017}. In these systems, entropically rich phases transcend conventional magnetic order and serve as a platform for realizing two-dimensional magnetic monopole gases (2DMG) \cite{Miao2020, Timsina2024}, which carry a non-zero net magnetic charge. To date, most investigations have focused on ground-state properties, leaving the thermodynamic behavior of frustrated phases at finite temperatures largely unexplored. Recently, Lu et al. \cite{Lu2024} reported strain-induced enhanced-entropy (EE) states in pyrochlore spin ice at low temperatures; however, the origin and how to engineer such EE states remain poorly understood.

In this study, we explore frustrated pyrochlore magnets $R_2$Ir$_2$O$_7$ ($R$ = Dy, Ho) by tuning the $d$-$f$ exchange interactions between $R^{3+}$ moments and neighboring $\text{Ir}^{4+}$ moments. Here, we systematically investigate the thermodynamic behavior and physical origin of two EE phases located at boundaries between three well-known stable phases: \enquote{2 spins in–2 spins out} (2I2O) spin ice phase, \enquote{3 spins in–1 spin out/1 spin in–3 spins out} (3I1O/1I3O) fragmented phase, and \enquote{all spins in–all spins out} (AIAO) antiferromagnetic state \cite{Lefrançois2017}. These EE phases are characterized by high entropy and high magnetic susceptibility, which are established only at finite temperatures. Our results indicate that these EE phases emerge through entropic stabilization due to free energy minimization, exhibiting distinct thermodynamic signatures from spin ice phase, fragmented phase, and antiferromagnetic phase, thus identifying a new class of EE magnetic states. Our findings reveal a foundation for engineering EE states in correlated magnetic systems, enabling the design of advanced materials with tunable memory and topological properties.

\section*{Results and discussions}
\section*{A. Tunable Magnetic Ground States and Frustration Levels}

\begin{figure*}
\begin{center}
\includegraphics[width=7in]{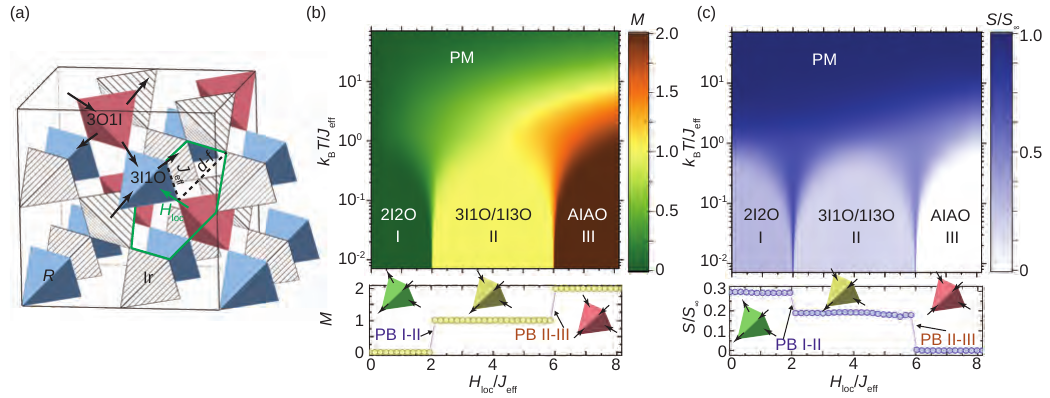}
\phantomsubcaption
\label{Fig1a}
\phantomsubcaption
\label{Fig1b}
\phantomsubcaption
\label{Fig1c}
\end{center}
\caption{Magnetic structure and frustration in a pyrochlore lattice. (a) Illustration of lattice and spin structure of pyrochlore $R_2$Ir$_2$O$_7$ magnets, where 3I1O/1I3O spin structures are sketched due to the intermediate strength of $d$-$f$ interaction as compared to $J_\text{eff}$ interaction. The tetrahedra associated with $R^{3+}$ moments are shown in blue (3I1O) and red (1I3O), while those associated with $\text{Ir}^{4+}$ moments are shaded white. (b) Magnetic order parameter $M$ defined as \cref{Eq:2}, as a function of $k_\text{B}T/J_\text{eff}$ and local static field $H_\text{loc}/J_\text{eff}$ (upper panel), and ground-state $M$ as a function of $H_\text{loc}/J_\text{eff}$ (lower panel). (c) Normalized entropy $S/S_\infty$ defined as \cref{Eq:3}, as a function of $k_\text{B}T/J_\text{eff}$ and $H_\text{loc}/J_\text{eff}$ (upper panel), and ground-state $S/S_\infty$ versus $H_\text{loc}/J_\text{eff}$ (lower panel). Three ground state phases, 2I2O, 3I1O/1I3O, AIAO, and the paramagnetic phase, are labeled as I, II, III, and PM, respectively. PB I-II and PB II-III stand for the phase boundaries between phases I/II and II/III, respectively.}
\label{Fig1}
\end{figure*}

In pyrochlore iridates $R_2$Ir$_2$O$_7$ ($R =$ Ho, Dy), both $R^{3+}$ and Ir$^{4+}$ ions form interconnected networks of corner-sharing tetrahedra, as illustrated in Fig. \ref{Fig1a}. The Ir$^{4+}$ sublattice forms AIAO antiferromagnetic order below $T_N \sim 130$ K, while the $R^{3+}$ moments typically order into a 2I2O spin ice configuration at $T \sim 1$ K. The disparity in these ordering temperatures results in the Ir$^{4+}$ moments acting as a static antiferromagnetic background field \( H_{\text{loc}}\) on the $R^{3+}$ sublattice via $d$–$f$ exchange coupling. The large ($\sim10 \mu_\text{B}$) $R^{3+}$ moments are strongly constrained by crystal-field and spin–orbit interactions to align along the local $\langle 111 \rangle$ axes, making a classical two-state Ising description an appropriate framework \cite{Bramwell2001,Lefrançois2017}. To quantitatively investigate the magnetic frustrations in pyrochlore iridates, we follow the classical nearest-neighbor pseudo-Ising Hamiltonian established for $R_2$Ir$_2$O$_7$ ($R =$ Ho, Dy) \cite{Lefrançois2017} and systematically vary $H_\text{loc}$:

\begin{equation} \label{Eq:1} \mathscr{H} = J_{\text{eff}} \sum_{\langle i,j \rangle} \sigma_i \sigma_j - \frac{1}{6} H_{\text{loc}} \sum_{\langle i,\alpha \rangle} \sigma_i \sigma_{\alpha}, \end{equation} 

where $\sigma_i = \pm 1$ represents the pseudo-Ising spins of $R^{3+}$ ions, and $\sigma_\alpha$ denotes the frozen AIAO-ordered spins of the Ir$^{4+}$ sublattice which is treated as a static background. The first term denotes the nearest-neighbor interactions among the $R^{3+}$ spins, while the second term indicates the interactions between $R^{3+}$ spins $\sigma_i$ and the static $\mathrm{Ir}^{4+}$ background spins $\sigma_\alpha$. We perform Monte Carlo simulations using the Metropolis algorithm \cite{Lefrançois2017, Miao2020, Timsina2024} by evolving the $R^{3+}$ spin configurations, with the Ir$^{4+}$ sublattice held fixed throughout.

We calculate the magnetic order parameter $M$ defined by:
\begin{equation} \label{Eq:2}
M = \left| \frac{1}{N} \sum_{i=1}^{N} \Delta_i q_i \right|,
\end{equation}

where $\Delta_i = +1$ ($-1$) for inward (outward) tetrahedra and $q_i$ is the magnetic charge index (all-in is 2, 3I1O is 1, 2I2O is 0, and so on) as a function of $k_\text{B}T/J_\text{eff}$ and $H_{\text{loc}}/J_{\text{eff}}$. The result demonstrates three distinct magnetic ground state phases consistent with previous studies \cite{Lefrançois2017}: the highly degenerate phase with 2I2O spin structure (phase I) for $H_{\text{loc}}/J_{\text{eff}}<2$, the fragmented phase with 3I1O/1I3O spin structure (phase II) for $2<H_{\text{loc}}/J_{\text{eff}}<6$, and the antiferromagnetic phase with AIAO spin structure (phase III) for $H_{\text{loc}}/J_{\text{eff}}>6$ as shown in Fig. \ref{Fig1b}. We also compute the normalized entropy per rare-earth moment $S(T)/S_\infty$ as a map of $k_\text{B}T/J_\text{eff}$ and $H_{\text{loc}}/J_{\text{eff}}$, where

\begin{equation} \label{Eq:3}
S(T) = S_\infty - \int_T^\infty \frac{C_v(T')}{T'} dT'
\end{equation}
and \( S_\infty = k_\text{B} \ln(2) \) is the infinite temperature entropy, as shown in Fig. \ref{Fig1c}. As expected, these phases exhibit distinct entropy behaviors that reflect their underlying frustration levels: the 2I2O phase retains significant nonzero spin ice Pauling entropy of $S =\frac{1}{2} k_\text{B} \ln(1.5)$ per spin \cite{Ramirez1999}, the 3I1O/1I3O phase carries intermediate entropy of $S \approx \frac{1}{2} k_\text{B} \ln(1.3)$ per spin \cite{Cathelin2020}, and the unfrustrated AIAO phase with zero entropy.

\section*{B. Thermodynamic Signatures Across Phase Boundaries}
\begin{figure}
\begin{center}
\includegraphics[width=3.5in]{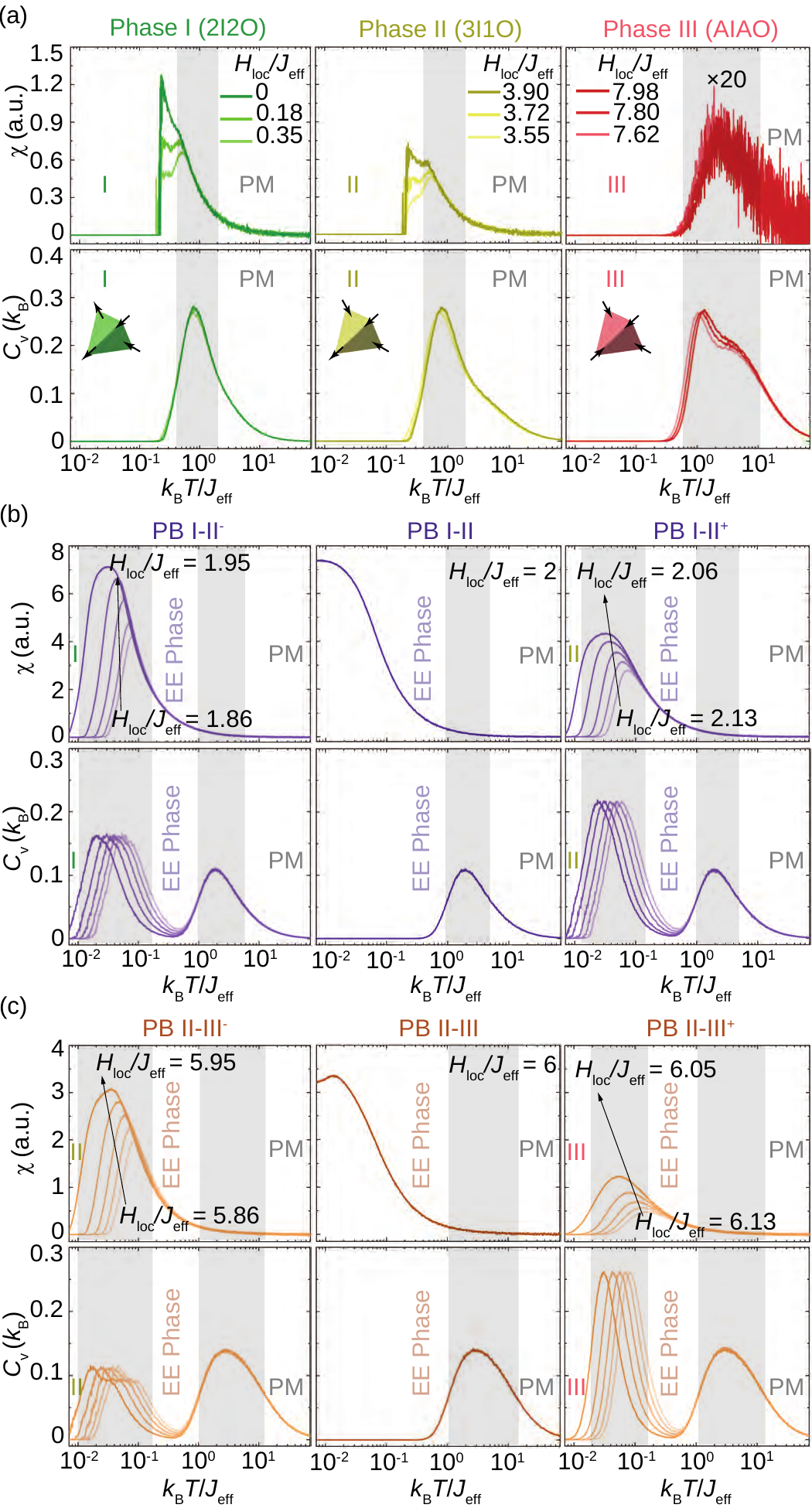}
\phantomsubcaption
\label{Fig2a}
\phantomsubcaption
\label{Fig2b}
\phantomsubcaption
\label{Fig2c}
\end{center}
\caption{Magnetic and thermodynamic response of pyrochlore $R_2$Ir$_2$O$_7$ magnets. (a) Magnetic susceptibility \( \chi \) (upper panel) and specific heat capacity \( C_v \) (lower panel) for phases I, II, and III as a function of normalized temperature $k_\text{B}T/J_\text{eff}$ for different $H_\text{loc}/J_\text{eff}$ values. (b) and (c) Magnetic susceptibility \( \chi \) (upper panel) and specific heat capacity \( C_v \) (lower panel) near the vicinity of PB I-II and PB II-III respectively, as a function of $k_\text{B}T/J_\text{eff}$ for various $H_\text{loc}/J_\text{eff}$ values. Shaded regions indicate the peak regions in the heat capacity, which are associated with phase transitions. We tentatively label the phase immediately below PM as an enhanced-entropy (EE) phase.}
\label{Fig2}
\end{figure}

To understand the thermodynamic behavior across different magnetic phases, we analyze the temperature dependence of magnetic susceptibility $\chi$ and specific heat capacity \( C_v \). In the stable phases, i.e. 2I2O (phase I), 3I1O/1I3O (phase II), and AIAO (phase III), each phase exhibits a broad, thermally activated peak in both $\chi$ and \( C_v \) in transition temperature regions, which are labeled with shaded regions as shown in Fig. \ref{Fig2a}. The approximate alignment of \( C_v \) and $\chi$ peaks indicate that spin ordering and thermal energy fluctuations occur cooperatively in the approach to magnetic orders.

However, this behavior changes significantly near the phase boundaries that separate these stable phases. In the vicinity of the phase boundary (PB) I–II, the temperature dependent behaviors in $\chi$ and \( C_v \) begin to decouple. For example, at PB I–II [Fig. \ref{Fig2b}], \( C_v \) exhibits a peak at around $k_\text{B}T/J_\text{eff}$ = 2.13 ($T$ = 3 K), indicating a phase transition, whereas $\chi$ increases monotonically without a peak toward the ground state. At vicinities of PB I-II on both sides, \( C_v \) exhibits two peaks indicating two phase transitions with a EE phase at finite temperature, whereas $\chi$ only exhibits one peak. Similar behavior is observed around PB II-III as shown in Fig. \ref{Fig2c}. We have varied the system sizes to confirm that the reported results are independent of finite-size effects (see Note 1 of Supplemental Material) \cite{SupplementalMaterial}. These behaviors are atypical for stable ground states and can be attributed to free spins mimicking thermal fluctuations, as detailed in Note 2 of Supplemental Material \cite{SupplementalMaterial}. The appearance of such anomalies further motivates a broader exploration of the temperature–field phase space, as discussed below.

\section*{C. Demonstration of EE Phases}
\begin{figure*}
\begin{center}
\includegraphics[width=7in]{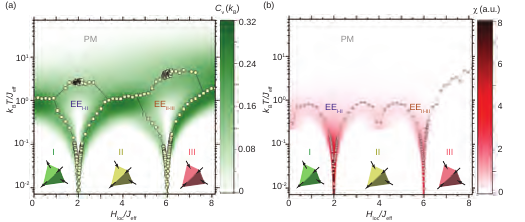}
\phantomsubcaption
\label{Fig3a}
\phantomsubcaption
\label{Fig3b}
\end{center}
\caption{Magnetic and thermodynamic response phase diagrams. (a) Specific heat capacity \( C_v \) and (b) Magnetic susceptibility \( \chi \) as a map of $k_\text{B}T/J_\text{eff}$ and $H_\text{loc}/J_\text{eff}$. Small circles filled with colors represent the peak positions of \( \chi \) and \( C_v \) as a function of $k_\text{B}T/J_\text{eff}$ and $H_\text{loc}/J_\text{eff}$. Phases I (2I2O), II (3I1O/1I3O), and III (AIAO) are labeled. Two EE phases are observed between phases I/II and II/III, as defined in the \( C_v \) map, and are labeled as $\text{EE}_{\text{I-II}}$ and $\text{EE}_{\text{II-III}}$, respectively.}
\label{Fig3}
\end{figure*}

To explore the thermodynamic landscape beyond the stable phases at the ground state, we map the magnetic susceptibility $\chi$ and specific heat $C_\text{v}$ over a broad range of $k_\text{B}T/J_\text{eff}$ and local static field $H_\text{loc}/J_\text{eff}$. The specific heat $C_\text{v}$ exhibits one peak near the center of the stable phases, i.e.$H_\text{loc}/J_\text{eff}=2,4,8$, whereas around the phase boundaries it bifurcates and exhibits two peaks. Between those peaks, $C_\text{v}$ has a small value close to 0, indicating the existence of new phases, as shown in Fig.~\ref{Fig3a}. Since they exist at finite temperatures and at the vicinity of phase boundaries, we tentatively name them $\text{EE}_\text{I-II}$ and $\text{EE}_\text{II-III}$, where \enquote{EE} stands for EE phases. These phases are consistent with previously reported step-like behavior in the magentic order parameter at finite temperature in the same system \cite{Lefrançois2017}. The presence of such EE states is further supported by the behavior of the magnetic susceptibility $\chi$, which remains significantly elevated across these regions. Unlike in ground-state-stable phases, where $\chi$ typically decreases to zero from the transition temperature into the ground state, here it maintains a high value over an extended temperature range, as shown in Fig.~\ref{Fig3b}. This persistent response indicates enhanced magnetic fluctuations due to free spins that change configuration between two adjacent stable phases.

\section*{D. Magnetic Structure and Thermodynamic Stability of EE Phases}
\begin{figure}
\begin{center}
\includegraphics[width=3.5in]{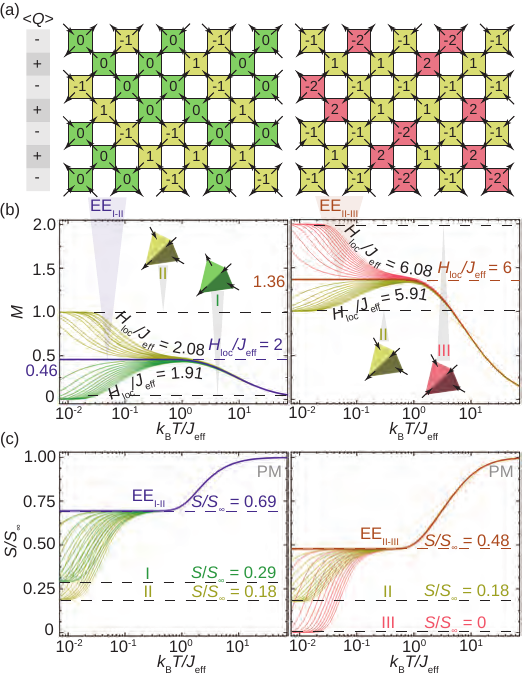}
\phantomsubcaption
\label{Fig4a}
\phantomsubcaption
\label{Fig4b}
\phantomsubcaption
\label{Fig4c}
\end{center}
\caption{Magnetic frustration and entropy of EE phases. (a) 2D illustration of EE phases $\text{EE}_{\text{I-II}}$ (left panel) and $\text{EE}_{\text{II-III}}$ (right panel). Tetrahedral sites with magnetic charge $Q = 0$ (phase I), $Q = \pm q_i$ (phase II), and $Q = \pm 2q_i$ (phase III) are shown in green, yellow, and light red, respectively. (b) $M$ as a function of $k_\text{B}T/J_\text{eff}$ for $\text{EE}_{\text{I-II}}$ (left panel) and $\text{EE}_{\text{II-III}}$ (right panel). (c) $S/S_\infty$ as a function of $k_\text{B}T/J_\text{eff}$ for $\text{EE}_{\text{I-II}}$ (left panel) and $\text{M}_\text{II-III}$ (right panel). Observables are shown over the range of $H_\text{loc}/J_\text{eff}$ with interval 0.007: from $H_\text{loc}/J_\text{eff} = 1.91$ to $2.08$ for $\text{EE}_{\text{I-II}}$, and from $5.91$ to $6.08$ for $\text{EE}_{\text{II-III}}$.}
\label{Fig4}
\end{figure}

To further understand the magnetic structure, stability, and disorderness of EE phases \( \text{EE}_{\text{I--II}} \) and \( \text{EE}_{\text{II--III}} \), we studied the behavior of the magnetic order parameter \( M \) and normalized entropy $S/S_\infty$ within narrow intervals of the local static field near the phase boundaries. At PB I-II with $H_{\text{loc}}/J_{\text{eff}}=2$, the magnetic order parameter $M$ rises and saturates at 0.46 at the ground state, indicating that $r_\text{I}\sim54\%$ of tetrahedra are 2I2O and $r_\text{II}\sim46\%$ of tetrahedra are 3I1O/1I3O. The spin structure and tetrahedral distribution are sketched in Fig.~\ref{Fig4a} (left panel). The positive and negative magnetic charged tetrahedra are located in alternating layers. We have performed Fourier transformation and did not observe any other ordering of these tetrahedra. Indeed, the ground state entropy of this EE state is as high as $S \approx 0.69S_\infty\approx \frac{1}{2} k_\text{B}\ln(2.6)$ per spin as shown in Fig.~\ref{Fig4c} (left panel), consistent with the absence of magnetic order. Close to the ground state, even small shifts in \( H_{\text{loc}}/J_{\text{eff}} \) result in sharp changes in \( M \) and $S/S_\infty$, reflecting a high sensitivity to the local static field as shown in Figs.~\ref{Fig4b} and ~\ref{Fig4c} (left panel) respectively. However, as temperature increases to $k_\text{B}T/J_\text{eff}$ = 0.71 (\( T \sim 1\,\text{K} \)), the system converges toward a common thermodynamic behavior with well-defined \( M \) and $S/S_\infty$. These behaviors indicate that the $\text{EE}_{\text{I--II}}$ phase is highly disordered and not stable at the ground state but stable at elevated temperatures. The step-like behavior of the order parameter is consistent with previous Monte Carlo simulation results \cite{Lefrançois2017}, and, together with the corresponding step-like behavior in entropy, provides evidence for the existence of the EE phase. The $\text{EE}_{\text{II--III}}$ phase exhibits a magnetic order parameter $M$ of 1.36, and a ground state entropy of $S \approx 0.48S_\infty\approx \frac{1}{2} k_\text{B} \ln(1.94)$ per spin. The magnetic structure should have $r_\text{II}\sim64\%$ percent of 3I1O/1I3O tetrahedra and $r_\text{III}\sim36\%$ of AIAO tetrahedra indicated in Fig.~\ref{Fig4a} (right panel). Similar thermodynamic behavior is observed around the $\text{EE}_{\text{II--III}}$ phase as shown in Figs.~\ref{Fig4b} and \ref{Fig4c} (right panel). We compute the uncertainties for the order parameter $M$ and normalized entropy $S/S_{\infty}$, which are very small, typically of order $10^{-4}$ and $5\times10^{-4}$, respectively (see Note 3 of Supplemental Material \cite{SupplementalMaterial} see also Ref. \cite{Ambegaokar2010} therein for a detailed error analysis).

The EE phase tetrahedra fractions and entropy can be understood by entropy maximization. Take EE$_\text{I-II}$ phase for example, the entropy can be naively viewed as a combination of the phase I (2I2O) configurational entropy $s_\text{I}=0.5k_\text{B}\ln(1.5)$ and the phase II (3I1O) configurational entropy $s_\text{II}=0.5k_\text{B}\ln(1.3)$, weighted by their respective volume fractions $r_\text{I}$ and $1-r_\text{I}$, together with an additional entropy term from the random mixing of tetrahedra. Accordingly, the entropy per spin can be expressed as
\[
S_\text{I-II} = r_\text{I}s_\text{I} + (1-r_\text{I})s_\text{II} + 0.5k_\text{B}\big[-r_\text{I}\ln r_\text{I} - (1-r_\text{I})\ln(1-r_\text{I})\big].
\]
At the phase I–II boundary, where both phases are energetically degenerate, the volume fraction ratio is determined by maximizing $S_\text{I-II}$. Solving $\text{d}S_\text{I-II}/\text{d}r_\text{I}=0$ yields $r_\text{I}=53.6\%$, $r_\text{II}=46.4\%$, and $S_\text{I-II}=74.0\%S_\infty$. Similarly, applying the same entropy model to the EE$_\text{II-III}$ phase gives $r_\text{II}=56.5\%$, $r_\text{III}=43.5\%$, and $S_\text{II-III}=60.0\%S_\infty$. Compared to the Monte Carlo simulation results in Fig.~\ref{Fig4}, these estimates are qualitatively consistent but show some quantitative discrepancies. Such discrepancies arise because this simplified entropy model does not account for additional geometric constraints that emerge when different types of tetrahedra coexist and share common spins, causing their configurational entropy contributions to deviate from the bulk values.

\section*{E. Entropic origin of EE phases}
\begin{figure}
\begin{center}
\includegraphics[width=3.5in]{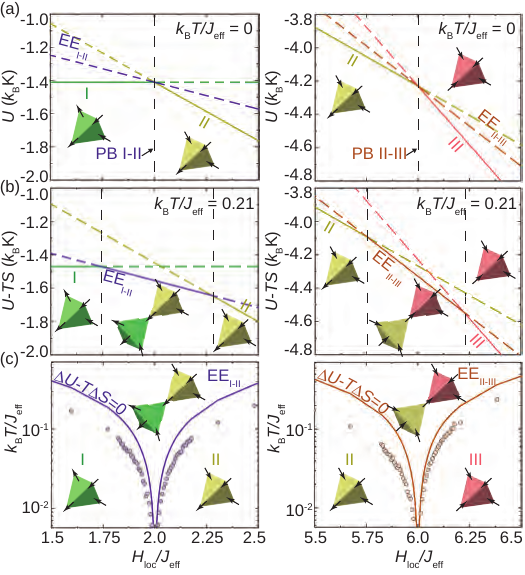}
\phantomsubcaption
\label{Fig5a}
\phantomsubcaption
\label{Fig5b}
\phantomsubcaption
\label{Fig5c}
\end{center}
\caption{Free energy mechanism for entropic stabilization of EE phases. (a) Internal energy \( U \) and (b) Free energy \( F = U - TS \) as a function of static background field \( H_{\text{loc}}/J_{\text{eff}} \), shown for regions surrounding the stable phase boundaries PB I--II (left panel) and PB II--III (right panel). Lowest and higher values of each \( H_{\text{loc}}/J_{\text{eff}} \) parameter are represented by solid and dashed lines, respectively. (c) Comparison between predicted and simulated EE phase boundaries. The solid lines were obtained from the equilibrium of free energy across the phase boundary as described in \cref{Eq:4}. Small circles filled with color represent the observed peak positions of \( C_v \) as a function of $k_\text{B}T/J_\text{eff}$ and \( H_{\text{loc}}/J_{\text{eff}} \) near to phase transitions.}
\label{Fig5}
\end{figure}

To understand the thermodynamic origin of EE phases, we analyze the energy, entropy, and free energy landscape of the system. These phases cannot be stabilized by minimization of internal energy alone, but emerge at finite temperatures due to the competition between internal energy and entropy, governed by the free energy expression $F=U-TS$. We begin by examining the energy landscape at zero temperature. Figure~\ref{Fig5a} shows the internal energy $U$ as a function of $H_{\text{loc}}/J_{\text{eff}}$ for the three stable phases, 2I2O (I), 3I1O/1I3O (II), and AIAO (III), as well as the two EE phases, $\text{EE}_{\text{I--II}}$ and $\text{EE}_{\text{II--III}}$. At $k_\text{B}T/J_\text{eff} = 0$, the system favors the configuration with the lowest energy: Phase I dominates for $H_{\text{loc}}/J_{\text{eff}}<2$, Phase II for $2<H_{\text{loc}}/J_{\text{eff}}<6$, and Phase III for $H_{\text{loc}}/J_{\text{eff}}>6$. The EE phases, while higher in energy, are close in magnitude near the phase boundaries, suggesting the possibility of competition once entropy is considered. To probe this entropic stabilization, we evaluate the free energy $F=U-TS$ at $k_\text{B}T/J_\text{eff} = 0.21$ ($T$ = 0.3 K), as shown in Fig.~\ref{Fig5b}. At this elevated temperature, EE phases now attain the lowest free energy in narrow windows near the respective phase boundaries, demonstrating that these phases are stabilized by entropy. To further test this picture, we estimate the boundaries of the EE regions by applying the free energy equilibrium condition
\begin{equation} \label{Eq:4}
\Delta F=\Delta U - T\Delta S=0
\end{equation}
The resulting contours from \cref{Eq:4} are plotted as solid lines in Fig.~\ref{Fig5c} and compared to the actual simulated phase boundaries extracted from heat capacity peaks (colored circles). The predicted trends agree qualitatively, capturing the cusp-like shape of the boundary near both EE phases, though the estimated crossover temperature is slightly higher than that observed in simulation. This discrepancy can be attributed to two simplifying assumptions in our analysis. First, the boundary estimation uses ground-state values of $U$ and $S$, neglecting their temperature dependence. Second, the heat capacity peaks reflect a characteristic crossover temperature in the simulation, whereas our analysis implicitly assumes a sharp first-order transition. In reality, the transitions into EE phases are gradual and fluctuation-driven, not symmetry-breaking. The emergence of EE phases is a direct consequence of entropy-driven free energy minimization in regions of high frustration and near-degenerate energy scales.

\section*{Discussion}
A promising route to engineer these EE phases is using an external magnetic field. By lifting the degeneracy among spin textures, the magnetic field reshapes the entropy landscape and allows selective tuning of EE phase boundaries. This tunability offers a practical route for field control in magnetic states.

An important extension of this work is to explore the role of quantum fluctuations. Our study has focused on classical pyrochlores with $R=$ Dy and Ho, where the large $R^{3+}$ moments behave as Ising spins along the local $\langle111\rangle$ axes. In contrast, for $R=$ Er or Yb, the crystal-field ground-state doublets are strongly anisotropic in the XY plane with reduced moments, leading to enhanced quantum fluctuations and tunneling \cite{Savary2012,Gingras2014}. In such cases, entropy may still play a role in stabilizing intermediate states, but the character of the degeneracy and correlations could be qualitatively different, raising the possibility of distinct thermodynamic signatures and quantum phases.

A natural next step is to investigate how the predicted EE phases can be probed experimentally. Thermodynamic probes such as specific heat and entropy measurements provide direct access to the characteristic entropy plateaus \cite{Ramirez1999}, while neutron scattering can resolve the associated spin correlations and dynamics \cite{Morris2009,Lefrançois2017}. Magnetic susceptibility would further reveal the field response predicted in this work \cite{Ramirez1999,Morris2009,Lefrançois2017}. To systematically explore the $H_{\mathrm{loc}}$ parameter space, one may employ chemical tuning, such as substituting titanium for iridium, or physical tuning through hydrostatic pressure. While chemical substitution offers a broad tuning range, it inevitably introduces disorder, whereas hydrostatic pressure modifies the lattice environment without adding extrinsic defects. Experimentally, EE states are expected to manifest as double anomalies in specific heat with temperature, together with a decoupling between thermodynamic peaks and magnetic susceptibility features.

The physics of EE phases studied in this work can also be applied to other magnetic systems with frustration and competing interactions, such as quantum spin liquids, artificial spin ice, and quantum spin ice. In these systems, metastability and entropy-driven phenomena may underlie transitions and emergent excitations, opening a gate into the discovery of exotic emergent quantum phases.

More broadly, the ability to stabilize and engineer EE phases has implications that reach beyond magnetism. In condensed matter physics, EE states often act as precursors to exotic phenomena—such as topological phases or non-equilibrium steady states \cite{Birch2021, Honda2020, Wheeler2025, Oike2016}. In materials science, EE states corresponds functionalities in memory devices \cite{Jo2022, Ravnik2021}, neuromorphic computing \cite{Markovic2020, Kudithipudi2025, Schuman2022, Yik2025}, and stochastic information encoding \cite{Ravichandran2024, Zhang2021}.

\section*{Conclusion}

In this study, we investigated the thermodynamic phase behavior of frustrated pyrochlore magnets \( R_2 \)Ir\(_2\)O\(_7\) (\( R = \) Dy, Ho) using Monte Carlo simulations by tuning the strength of the local static background field \( H_{\text{loc}}/J_{\text{eff}} \). Beyond the three known magnetic ground states, frustrated spin ice 2-in–2-out (2I2O) phase, frustrated fragmented 3-in–1-out/1-in–3-out (3I1O/1I3O) phase, and antiferromagnetic all-in–all-out (AIAO) phase, we systematically studied the thermodynamic behavior and physical origin of two enhanced-entropy (EE) phases ($\text{EE}_{\text{I-II}}$ and $\text{EE}_{\text{II-III}}$) that appear near the phase boundaries at finite temperatures. These EE states exhibit high magnetic susceptibility, high entropy, and spatially mixed spin configurations of neighboring phases, consistent with enhanced magnetic fluctuations and configurational disorder. Our findings suggest that these states do not result from conventional energetic ordering but are stabilized by entropy through the minimization of free energy in regions of frustrated competition. Unlike typical ordered phases, they remain disordered yet thermodynamically robust across extended temperature and \( H_{\text{loc}}/J_{\text{eff}} \) ranges. This demonstrates the pivotal role of entropy in governing phase stability in frustrated systems and opens a new avenue for exploring entropic phases in correlated magnets.

\section*{Methods}

We performed Monte Carlo simulations based on the Ising model on frustrated pyrochlore \( R_2 \)Ir\(_2\)O\(_7\) (\( R = \) Dy, Ho) magnets. These simulations explore accessible spin configurations to determine the thermodynamically probable magnetic states. The model includes nearest-neighbor exchange interactions between the pseudo-Ising moments of the rare-earth ions (\( R^{3+} \)), represented as classical binary spins constrained along local \(\langle 111 \rangle\) directions. Simulations were carried out on a \( 4 \times 4 \times 8 \) pyrochlore lattice with periodic boundary conditions. The spin dynamics were evolved using a hybrid Monte Carlo algorithm that combines single-spin-flip (SSF) Metropolis updates with a worm-loop algorithm (WLA). This hybrid approach ensures ergodicity and enhances phase-space exploration, particularly at low temperatures, preventing the system from becoming trapped in local minima \cite{Lefrançois2017}.

We use a two-step sampling procedure. The first step is thermal equilibrium, where we perform 2000 cycles of SSF/WLA hybrid updates to allow the system to reach thermal equilibrium. Each cycle includes 1,000 Monte Carlo steps. The convergence of system energy indicating thermal equilibrium are monitored and confirmed (see Supplemental Material note 4 \cite{SupplementalMaterial}). Following equilibration, we perform 4000 additional cycles. In each cycle there are also 1,000 Monte Carlo steps followed by measurements physical observables including internal energy, monopole density, and other quantities of interest. A total of approximately 1000 temperature steps were carried out. Thermodynamic observables were extracted from the equilibrated spin configurations. The internal energy \( U \) per spin was computed using the nearest-neighbor Ising Hamiltonian, and the specific heat capacity \( C_v \) was calculated from the temperature derivative of energy. The order parameter \( M \) was evaluated using the expression in \cref{Eq:2}, and the entropy per spin \( S(T) \) was obtained via \cref{Eq:3}. Magnetic susceptibility \( \chi \) was determined through a linear response method by applying a small static magnetic field $B_0$ to the system. Specifically, the simulation proceeds in two stages: (1) the system is equilibrated with and without the magnetic field $B_0$, and (2) the change in magnetization $\Delta m$ due to the application of the magnetic field divided by $B_0$. The susceptibility is then computed as $\chi = \Delta m/B_0$.

All simulations were implemented in C++. Additional implementation strategies and technical background are discussed in prior work \cite{Miao2020,Timsina2024}.

\section*{Data availability}
The data are not publicly available. The data are available from the authors upon reasonable request.

\section*{Acknowledgements} This work was supported by the National Science Foundation's Partnerships for Research and Education in Materials with award number 2423992. It was performed, in part, at the Center for Integrated Nanotechnologies, an Office of Science User Facility operated for the U.S. Department of Energy (DOE) Office of Science by Los Alamos National Laboratory (Contract 89233218CNA000001), in partnership with the LANL Institutional Computing Program for computational resources. We would like to acknowledge Dr. Christopher Lane, staff scientist at Los Alamos National Laboratory, for his assistance with the use of LANL Institutional Computing resources.

\section*{Additional information}
Correspondence and requests for materials should be addressed to L. M.

\section*{Competing financial interests}
The authors declare no competing interests.

\end{document}